\documentclass[12pt,preprint]{aastex}

\usepackage{graphicx}

\shorttitle{Active Region Modeling}
\shortauthors{Warren \& Winebarger}

\slugcomment{\texttt{Submitted to ApJ 01-SEP-2006}}

\begin{document}

%% ----------------------------------------------------------------------
%% --- TITLE PAGE -------------------------------------------------------
%% ----------------------------------------------------------------------

\title{Static and Dynamic Modeling of a Solar Active Region. I: Soft
X-Ray Emission}

\author{Harry P. Warren}
\affil{E. O. Hulburt Center for Space Research, Code 7673HW, Naval
  Research Laboratory, Washington, DC 20375}
\email{hwarren@nrl.navy.mil}
\and
\author{Amy R. Winebarger} 
\affil{Department of Physics, Alabama A\&M University, 4900 Meridian
Street Normal, AL 35762}
\email{winebarger@physics.aamu.edu}

%% ----------------------------------------------------------------------
%% --- ABSTRACT ---------------------------------------------------------
%% ----------------------------------------------------------------------

\begin{abstract}
 Recent simulations of solar active regions have shown that it is
 possible to reproduce both the total intensity and the general
 morphology of the high temperature emission observed at soft X-ray
 wavelengths using static heating models. There is ample observational
 evidence, however, that the solar corona is highly variable,
 indicating a significant role for dynamical processes in coronal
 heating. Because they are computationally demanding, full
 hydrodynamic simulations of solar active regions have not been
 considered previously. In this paper we make first application of an
 impulsive heating model to the simulation of an entire active region,
 AR8156 observed on 1998 February 16. We model this region by coupling
 potential field extrapolations to full solutions of the
 time-dependent hydrodynamic loop equations. To make the problem more
 tractable we begin with a static heating model that reproduces the
 emission observed in 4 different \textit{Yohkoh} Soft X-Ray Telescope
 (SXT) filters and consider dynamical heating scenarios that yield
 time-averaged SXT intensities that are consistent with the static
 case. We find that it is possible to reproduce the total observed
 soft X-ray emission in all of the SXT filters with a dynamical
 heating model, indicating that nanoflare heating is consistent with
 the observational properties of the high temperature solar corona.
\end{abstract}


%% ----------------------------------------------------------------------
%% --- INTRODUCTION -----------------------------------------------------
%% ----------------------------------------------------------------------

\section{Introduction}

 Understanding how the Sun's corona is heated to high temperatures
 remains one of the most significant challenges in solar physics.
 Unfortunately, the complexity of the solar atmosphere, with its many
 disparate spatial and temporal scales, makes it impossible to
 represent with a single, all encompassing model. Instead we need to
 break the problem up into smaller, more manageable pieces (e.g., see
 the recent review by \citealt{klimchuk2006}). For example, kinetic
 theory or generalized MHD is used to describe the microphysics of the
 energy release process. Ideal and resistive MHD are used to study the
 evolution of coronal magnetic fields and the conditions that give
 rise to energy release. Finally, one dimensional hydrodynamical
 modeling is employed to calculate the response of the solar
 atmosphere to the release of energy.

 This last step is a critical one in the process of understanding
 coronal emission because it links theoretical models with solar
 observations. Even here, however, most previous work has focused on
 modeling small pieces of the Sun, such as individual loops
 \citep[e.g.,][]{aschwanden2001b,reale2000}. Though understanding the
 heating in individual structures is an important first step, it has
 been difficult to apply this information to constrain the properties
 of the global coronal heating mechanism.

 Recent advances in high performance computing have made it possible
 to simulate large regions of the corona, at least with static heating
 models. \cite{schrijver2004}, for example, have coupled potential
 field source-surface models of the coronal magnetic field with
 parametric fits to solutions of the hydrostatic loop equations to
 calculate visualizations of the full Sun. Comparisons between the
 simulation results and full-disk solar images indicate that the
 energy flux ($F_H$) into a corona loop scales as $B_F/L$, where $B_F$
 is the foot point field strength and $L$ is the loop
 length. \cite{schrijver2005b} also find that this form for the
 heating flux is consistent with the flux-luminosity relationship
 derived from X-ray observations of other cool dwarf stars
 \cite{schrijver2005b}.

 \cite{warren2006b} have performed similar simulations for 26 solar
 active regions using potential field extrapolations and full
 solutions to the hydrostatic loop equations. These simulation results
 indicate that the observed emission is consistent with a volumetric
 heating rate ($\epsilon_S$) that scales as $\bar{B}/L$, where
 $\bar{B}$ is the field strength averaged along the field line. In the
 sample of active regions used in that study $\bar{B}\sim B_F/L$ so
 that $F_H\sim \epsilon_S L \sim \bar{B}\sim B_F/L$, and this form for
 the volumetric heating rate is consistent with the energy flux
 determined by \cite{schrijver2004}.

 In these previous studies it was possible to use static heating
 models to reproduce the high temperature emission observed at soft
 X-ray wavelengths, but not the lower temperature emission typically
 observed in the EUV. The static models are not able to account for
 the EUV loops evident in the solar images. Recent work has shown that
 the active region loops observed at these lower temperatures are
 often evolving \citep{urra2006,winebarger2003b}. Simulation results
 suggest that these loops can be understood using dynamical models
 where the loops are heated impulsively and are cooling
 \citep[e.g.,][]{spadaro2003,warren2003}. Furthermore, spectrally
 resolved observations have indicated pervasive red shifts in active
 regions at upper transition region temperatures
 \citep[e.g.,][]{winebarger2002}, suggesting that much of the plasma
 in solar active regions near 1\,MK has been heated to higher
 temperatures and is cooling.  Finally, \cite{warren2006b} found that
 static heating in loops with constant cross section yields footpoint
 emission that is much brighter than what is observed. This suggests
 that static heating models may not be consistent with the
 observations, even in the central cores of active regions.

 The need for exploring dynamical heating models of the solar corona
 is clear, but there are a number of problems that make this difficult
 in practice. One problem is the many free parameters possible in
 parameterizations of impulsive heating models. In addition to the
 magnitude and spatial location of the heating, it is possible to vary
 the temporal envelope and repetition rate of the heating
 \citep[e.g.,][]{patsourakos2006,testa2005}. Furthermore, dynamical
 solutions to the hydrodynamic loop equations are much more
 computationally intensive to calculate than static solutions,
 limiting our ability to explore parameter space.

 In this paper we explore the application of impulsive heating models
 to the high temperature emission observed in active region 8156 on
 1998 February 16. To make the problem more tractable we begin with a
 static heating model that reproduces the emission observed in 4
 different \textit{Yohkoh} Soft X-Ray Telescope (SXT) filters and look
 for dynamical models that yield time-averaged SXT intensities that
 are in agreement with those computed from the static
 solutions. Relating the time-averaged intensities derived from the
 full dynamical solutions to the observed intensities is based on the
 idea that the emission from a single feature results from the
 superposition of even finer, dynamical structures that are in various
 stages of heating and cooling. This idea is similar to the nanoflare
 model of coronal heating
 \citep[e.g.,][]{parker1983,cargill1994}. Other nanoflare heating
 scenarios are possible, such as heating events on larger scale
 threads that are distributed randomly in space and time, but are not
 considered here. We find that it is possible to construct a dynamical
 heating model that reproduces the total soft X-ray emission in each
 SXT filter. This indicates that nanoflare heating is consistent with
 the observational properties of the high temperature corona.

 \section{Observations}

 Observations from SXT \citep{tsuneta1991} on \textit{Yohkoh} form the
 basis for this work. The SXT, which operated from late 1991 to late
 2001, was a grazing incidence telescope with a nominal spatial
 resolution of about 5\arcsec\ (2\farcs5 pixels). Temperature
 discrimination was achieved through the use of several focal plane
 filters. The SXT response extended from approximately 3\,\AA\ to
 approximately 40\,\AA\ and the instrument was sensitive to plasma
 above about 2\,MK.

 In addition to the SXT images we use full-Sun magnetograms taken with
 the MDI instrument \citep{scherrer1995} on \textit{SOHO} to provide
 information on the distribution of photospheric magnetic fields. The
 spatial resolution of the MDI magnetograms is comparable to the
 spatial resolution of EIT and SXT. In this study we use the synoptic
 MDI magnetograms which are taken every 96 minutes.

 To constrain the static heating model we require observations of an
 active region in multiple SXT filters. Observations in the thickest
 SXT analysis filters, the ``thick aluminum'' (Al12) and the
 ``beryllium'' (Be119), are crucial for this work. As we will show,
 observations in the thinner analysis filters, such as the ``thin
 aluminum'' (Al.1) and the ``sandwich'' (AlMg) filters, do not have
 the requisite temperature discrimination for this modeling. To
 identify candidate observations we made a list of all SXT
 partial-frame (as opposed to full disk) observations with
 observations in the Al.1, AlMg, Al12, and Be119 filters between the
 beginning of the \textit{Yohkoh} mission and the end of 2001. Since
 the potential field extrapolation is also important to this analysis,
 we required that the active lie within 400\arcsec\ of disk center.  

 We also use consider observations from the EIT
 \citep{delaboudiniere1995} on \textit{SOHO}. EIT is a normal
 incidence telescope that takes full-Sun images in four wavelength
 ranges, 304\,\AA\ (which is generally dominated by emission from
 \ion{He}{2}), 171\,\AA\ (\ion{Fe}{9} and \ion{Fe}{10}), 195\,\AA\
 (\ion{Fe}{12}), and 284\,\AA\ (\ion{Fe}{15}). EIT has a spatial
 resolution of 2\farcs6. Images in all four wavelengths are typically
 taken 4 times a day and these synoptic data are used in this study.

\begin{figure*}[t!]
 \centerline{%
 \includegraphics[angle=90,scale=0.85]{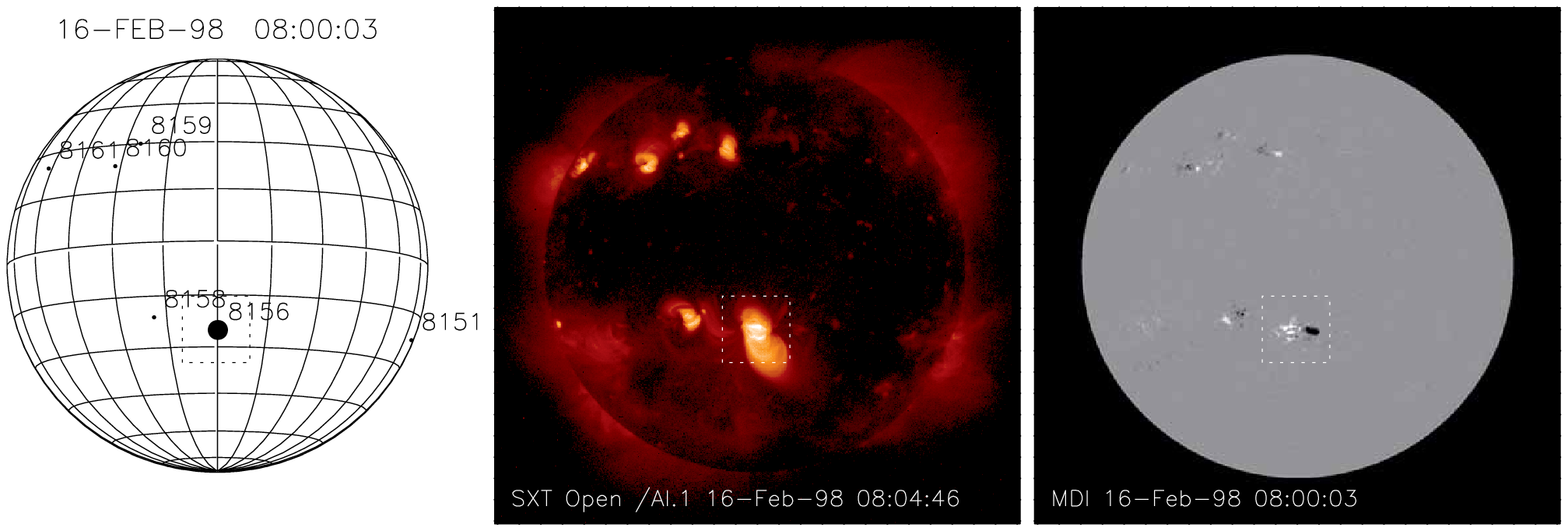}}
  \caption{Observations of AR8156 on 1996 February 16 near 08:00 UT in
  SXT (center) and MDI (right). The field of view of the SXT partial
  frame images is indicated by the box.}
 \label{fig:1}
 \end{figure*}

 From a visual inspection of the available data we selected
 observations of AR8156 taken 1998 February 16 near 8\,UT. This region
 is shown in full-disk SXT and MDI images in Figure~\ref{fig:1}. The
 region of interest observed in SXT, EIT, and MDI is shown in
 Figure~\ref{fig:2}. These images represent the observations taken
 closest to the MDI magnetogram. The total intensities in the SXT
 partial frame images for this region during the period beginning 1998
 February 15 23:30 UT and ending 1998 February 16 13:00 UT are
 generally within $\pm20$\% of the total intensities in these SXT
 images, indicating an absence of major flare activity during this
 time.

 \section{Static Modeling}

\begin{figure*}[t!]
 \centerline{%
 \includegraphics[angle=90,scale=0.85]{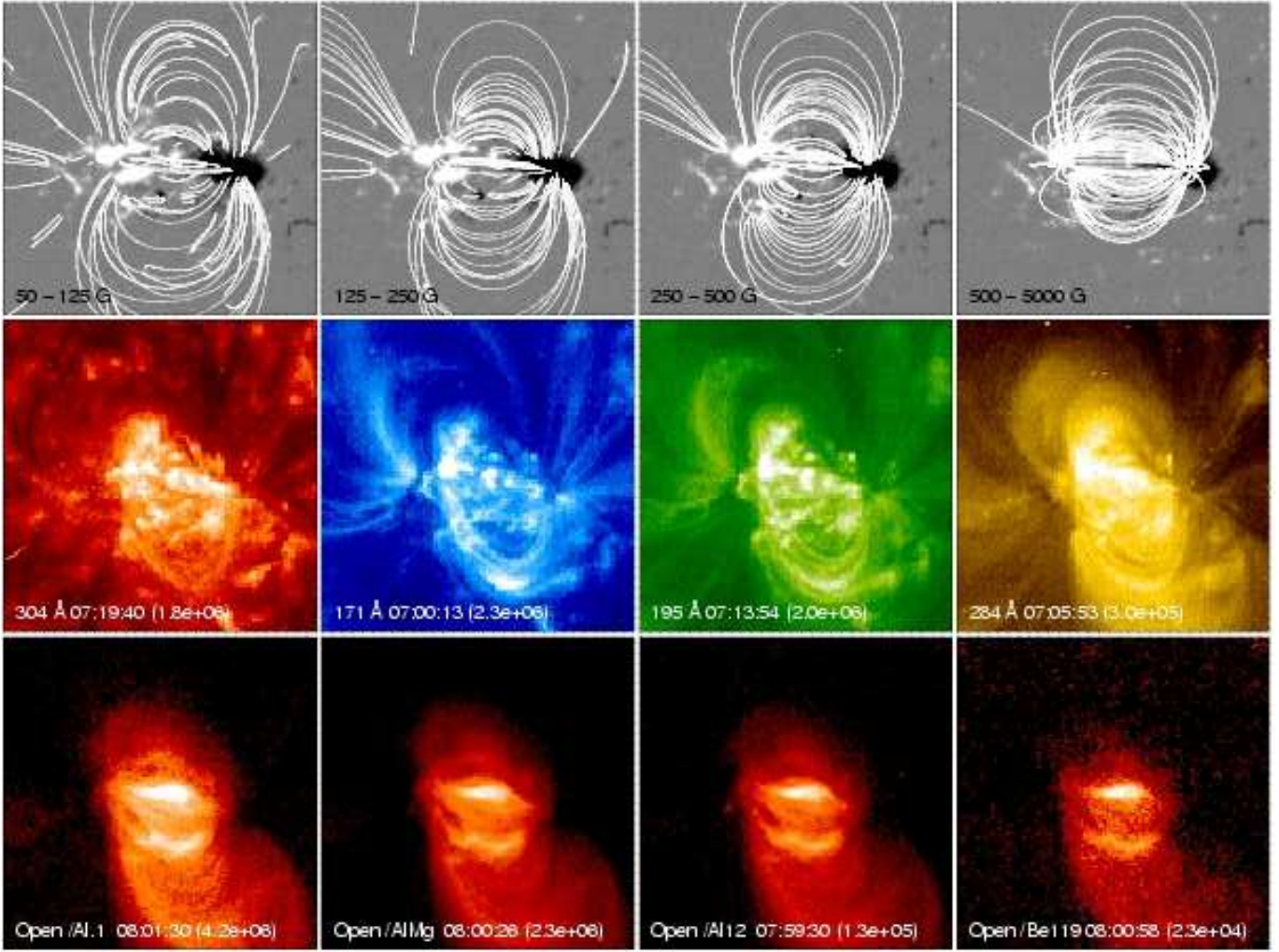}}
  \caption{(\textit{top panels}) Selected field lines from the
  potential field extrapolation of the MDI magnetogram taken at
  08:00:03 UT. (\textit{middle panels}) EIT synoptic images of AR8156
  in all four EIT wavelengths. (\textit{bottom panels}) SXT images in
  four filters.}
 \label{fig:2}
 \end{figure*}

 To model the topology of this active region we use a simple potential
 field extrapolation of the photospheric magnetic field. For each MDI
 pixel with a field strength greater than 50\,G we calculate a
 field line. Some representative field lines are shown in
 Figure~\ref{fig:2}.  It is clear that such a simple model does not
 fully reproduce the observed topology of the images. The long loops
 in the bottom half of the images, for example, are shifted relative
 to the field lines computed from the potential field
 extrapolation. However, as we argued previously \citep{warren2006b},
 our goal is not to reproduce the exact morphology of the active
 region. Rather, we are primarily interested in the more general
 properties of the active region emission, such as the total intensity
 or the distribution of intensities. The potential field extrapolation
 only serves to provide a realistic distribution of loop lengths.

 One subtlety with coupling a potential field extrapolation with
 solutions to the hydrostatic loop equations is the difference in
 boundary conditions. The field lines originate in the photosphere
 where the plasma temperature is approximately 4,000\,K. The boundary
 condition for the loop footpoints, however are typically set at
 10,000 or 20,000\,K in the numerical solutions to the hydrodynamic
 loop equations. Furthermore, studies of the topology of the quiet Sun
 have shown that a significant fraction of the field lines close at
 heights below 2.5\,Mm, a typical chromospheric height
 \citep{close2003}. To avoid these small scale loops we use the
 portion of the field line above 2.5\,Mm in the modeling and exclude
 all field lines that do not reach this height.

 For each of the 1956 field lines ultimately included in the
 simulation we calculate a solution to the hydrostatic loop equations
 using a numerical code written by Aad van Ballegooijen (e.g.,
 \citealt{hussain2002,schrijver2005}). Following our previous work,
 our volumetric heating function is assumed to be
 \begin{equation}
  \epsilon_S = \epsilon_0 \left(\frac{\bar{B}}{\bar{B}_0}\right)
  \left(\frac{L_0}{L}\right),
 \label{eq:heating}
 \end{equation}
 where $\bar{B}$ is the field strength averaged along the field line,
 and $L$ is the total loop length. We assume a constant cross section
 and a uniform distribution of heating along each loop. Note that the
 variation in gravity along the loop is determined from the geometry
 of the field line.

 The numerical solution to the hydrostatic loop equation provides the
 variation in the density, temperature, and velocity along the
 loop. The temperatures, densities, and loop geometry are then used to
 compute the expected response in the SXT and EIT filters. For our
 work we use the CHIANTI atomic database \citep[e.g.][]{dere1997} to
 compute the instrumental responses and the radiative losses used in
 the hydrostatic code (see \citealt{brooks2006} for a discussion of
 the instrumental responses and radiative losses).

 In our previous work the value for $\epsilon_0$ was chosen to be
 0.0492\,erg~cm$^{-3}$~s$^{-1}$ so that a ``typical'' field line
 ($\bar{B}=\bar{B}_0$ and $L=L_0$) had an apex temperature of
 $T_0=4$\,MK. We also found that for this value of $\epsilon_0$ a
 filling factor of about 10\% was needed to reproduce the SXT emission
 observed in the Al.1 or AlMg filters. In the absence of information
 from the hotter SXT filters the value for $\epsilon_0$ is poorly
 constrained. The values adopted for the parameters $\bar{B}_0$ and
 $L_0$ are 76\,G and 29\,Mm respectively.

\begin{figure*}[t!]
 \centerline{%
 \includegraphics[angle=90,scale=0.75]{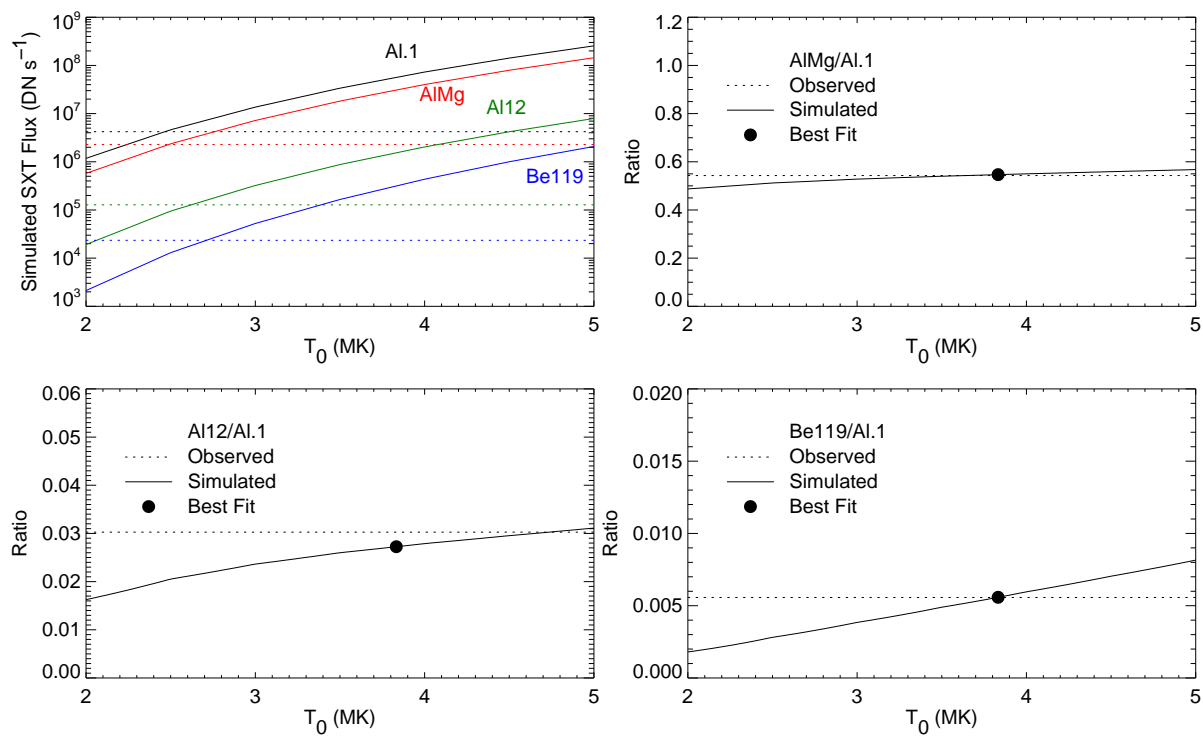}}
 \caption{Simulated and observed SXT intensities for AR8156. (\textit{top
 left panel}) The simulated intensities as a function of $T_0$ assuming a
 filling factor of unity. The observed intensities in each filter are
 indicated with the dotted lines. Simulations are calculated in steps
 of 0.5\,MK. (\textit{other panels}) Simulated filter ratios, which
 are independent of the filling factor, as a function of $T_0$. The
 observed filter ratios are indicated by the dotted lines. The best
 fit value for $T_0$ is also indicated on these plots.}
 \label{fig:3}
 \end{figure*}

 For this active region we have observations in the hotter SXT filters
 so we have performed active region simulations for a range of $T_0$
 (equivalently $\epsilon_0$) values. The resulting total intensities
 in each of the 4 filters as a function of $T_0$ are shown in
 Figure~\ref{fig:3}. It is clear from this figure that the static
 model cannot reproduce all of the SXT intensities for a filling
 factor of 1. For a filling factor of 1 the value of $T_0$ needed to
 reproduce the Al.1 intensity yields a Be119 intensity that is too
 low. Similarly, the value of $T_0$ that reproduces the Be119
 intensity for a filling factor of 1 yields Al.1 intensities that are
 too large. 

 By doing a least squares fit of the simulation results to the
 observations and varying both the value of $T_0$ and the filling
 factor we find that we can reproduce all of the SXT intensities to
 within 10\% for $T_0=3.8$\,MK and a filling factor of 7.6\%, values
 close to what we used in our previous work. These simulation results
 also highlight the importance of the SXT Al12 and Be119 filters in
 modeling the observations. The ratio between the Al.1 and AlMg
 filters is simply too shallow to be of any use in constraining the
 magnitude of the heating. The Al.1 to Be119 ratio, in contrast,
 varies by almost an order of magnitude as $T_0$ is varied from 2 to
 5\,MK.

 The total intensity represents the minimum level of agreement between
 the simulation and the observations. The distribution of the
 simulated intensities must also look similar to what is observed. To
 transform the 1D intensities into 3D intensities we assume that the
 intensity at any point in space is related to the intensity on the
 field line by
 \begin{equation}
  I(x,y,z) = kI(x_0,y_0,z_0)\exp\left[ -\frac{\Delta^2}{2\sigma_r^2}\right]
 \end{equation}
 where $\Delta^2 = (x-x_0)^2+(y-y_0)^2+(z-z_0)^2$ and $2.355\sigma_r$,
 the FWHM, is set equal to the assumed diameter of the flux tube. A
 normalization constant ($k$) is introduced so that the integrated
 intensity of over all space is equal to the intensity integrated
 along the field line. This approach for the visualization is based on
 the method used in \cite{karpen2001}. 

\begin{figure*}[t!]
 \centerline{%
 \includegraphics[angle=90,scale=0.80]{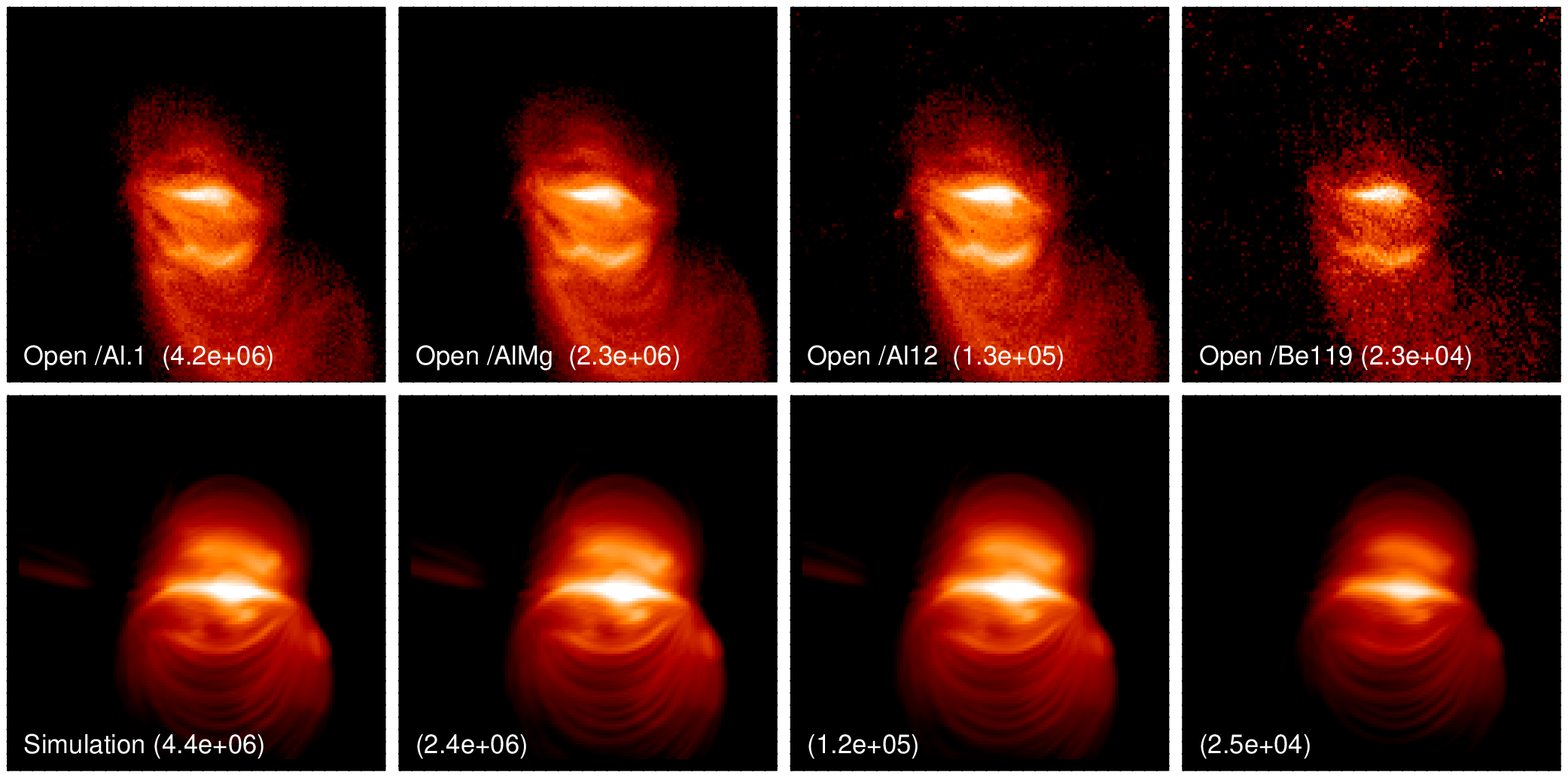}}
 \centerline{%
 \includegraphics[angle=90,scale=0.80]{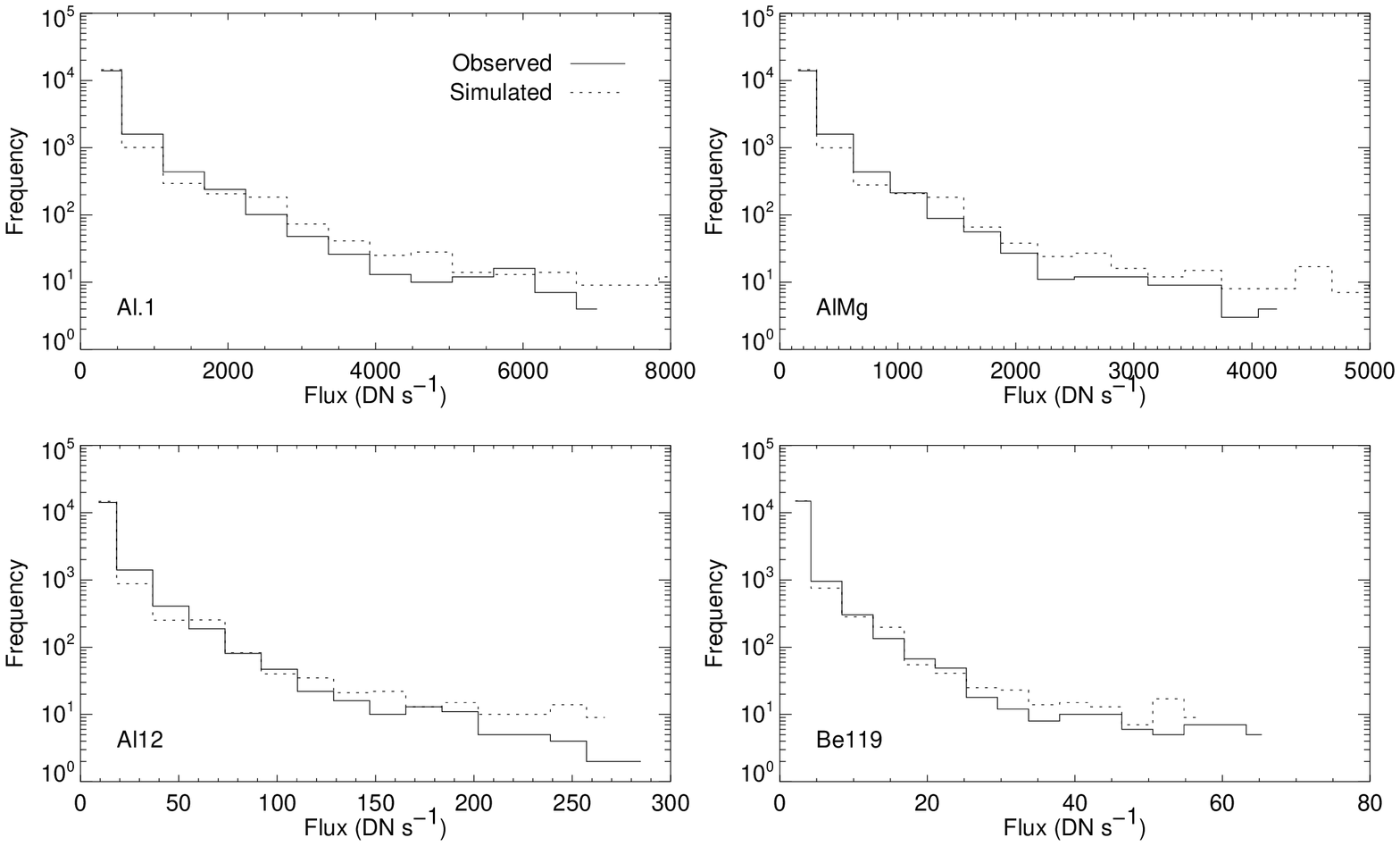}
 }
 \caption{Observed and simulated SXT emission for AR8156. (\textit{top
 panels}) Observed and simulated images. The calculated SXT images
 have been convolved with the SXT point spread function. The numbers
 in parentheses indicate the total intensity in the image. (\textit{bottom
 panels}) The intensity distributions for the observed and simulated
 images.}
 \label{fig:4}
 \end{figure*}

 The resulting simulated SXT images are shown in Figure~\ref{fig:4}
 where they are compared with the observations. The simulations
 clearly do a reasonable job reproducing these data, particularly in
 the core of the active region. At the periphery of the active region
 the simulation does not match either the morphology of the emission
 or its absolute magnitude exactly. The general impression, however,
 is that the model intensities are generally similar to the observed
 intensities outside of the active region core even if the morphology
 doesn't match exactly.

 One change that we have made from our previous methodology
 \citep{warren2006b} is to include all of the field lines with
 footpoint field strengths above 500\,G. These field lines have been
 largely excluded in previous work because sunspots are generally
 faint in soft X-rays images (see
 \citealt{golub1997,schrijver2004,fludra2003}). In these observations,
 however, the exclusion of the field lines rooted in strong field
 leads to small, but noticeable differences between the simulated and
 observed emission. As can be inferred from Figure~\ref{fig:2},
 excluding these field lines leads to an absence of emission on either
 side of the bright feature in the center of the active region. This
 suggests that the algorithm used to select which field lines are
 included in the simulation needs to be studied more carefully.

 The histograms of the intensities offer an additional point of
 comparison between the simulations and the observations. As shown in
 Figure~\ref{fig:4}, the distributions of the intensities are very
 similar in both cases, supporting the qualitative agreement between the
 visualizations and the actual solar images.

 \section{Dynamic Modeling}

 The principal difficulties with full hydrodynamic modeling of solar
 active regions are the many degrees of freedom available to
 parameterize the heating function and the computational difficulty of
 calculating the numerical solutions. For this exploratory study we
 make several simplifying assumptions. First, we consider dynamic
 simulations that are closely related to the static solutions. Since
 the static modeling of the SXT observations presented in the previous
 section adequately reproduces the total intensities, the distribution
 of intensities, and the general morphology of the images, it seems
 reasonable to consider dynamical heating that would reproduce the
 static solutions in some limit. Second, we utilize the time-averaged
 properties of these solutions in computing the simulated
 intensities. Our assumption is that the emission from a single field
 line in the static model actually results from the superposition of
 even finer, dynamical structures that are in various stages of
 heating and cooling. This is similar to the nanoflare picture of
 coronal heating \citep[e.g.,][]{parker1983,cargill1994}.  Finally, we
 will also make use of grids of solutions where we interpolate to
 determine the simulated intensities rather than computing solutions
 for each field line individually.

 In the static case we have used the average magnetic field strength
 and loop length to infer the volumetric heating rate ($\epsilon_s$)
 for each field line. For the dynamic case we consider volumetric
 heating rates of the form
 \begin{equation}
 \epsilon_D(t) = g(t)R\epsilon_S + \epsilon_B,
 \label{eq:heatingD}
 \end{equation}
 where $g(t)$ is a step or boxcar function envelope on the heating,
 $\epsilon_S$ is the static heating rate determined from
 Equation~\ref{eq:heating}, $R$ is a arbitrary scaling factor, and
 $\epsilon_B$ is a weak background heating rate that establishes a
 cool, tenuous equilibrium atmosphere in the loop. To solve the
 hydrodynamic loop equations numerically we use the NRL Solar Flux
 Tube Model (SOLFTM) code \cite[e.g.,][]{mariska1987,mariska1989}. 

 In the limit of an infinite heating window and $R=1$ the dynamic
 solutions would converge to the static solutions and all of the
 properties of the static simulation would be recovered. This is the
 primary motivation for our choice of the heating function given in
 Equation~\ref{eq:heatingD}. The good agreement between the
 observations and the static model suggest that the energetics of the
 static model are not far off.

 For $R=1$ and a finite duration to the heating we expect that the
 calculated SXT emission will generally be less than in the static
 case because it takes a finite time for chromospheric plasma to
 evaporate up into the loop. Thus simply truncating the heating will
 not produce acceptable results. If we increase the heating somewhat
 from the static case ($R>1$) and consider a finite duration we expect
 larger SXT intensities relative to the $R=1$, finite duration case
 since the evaporation will be faster and the temperatures will be
 higher with the increased heating. Since the time to fill the loop
 with plasma will depend on the sound crossing time ($\tau_s\sim
 L/c_s$, with $c_s$ the sound speed) the behavior of the dynamic
 solutions relative to the static solutions will also depend on loop
 length. For a finite duration to the heating the intensities in the
 dynamical simulations of the shorter loops will more closely resemble
 the results from the static solution.

\begin{figure*}[t!]
 \centerline{%
 \includegraphics[angle=90,scale=0.70]{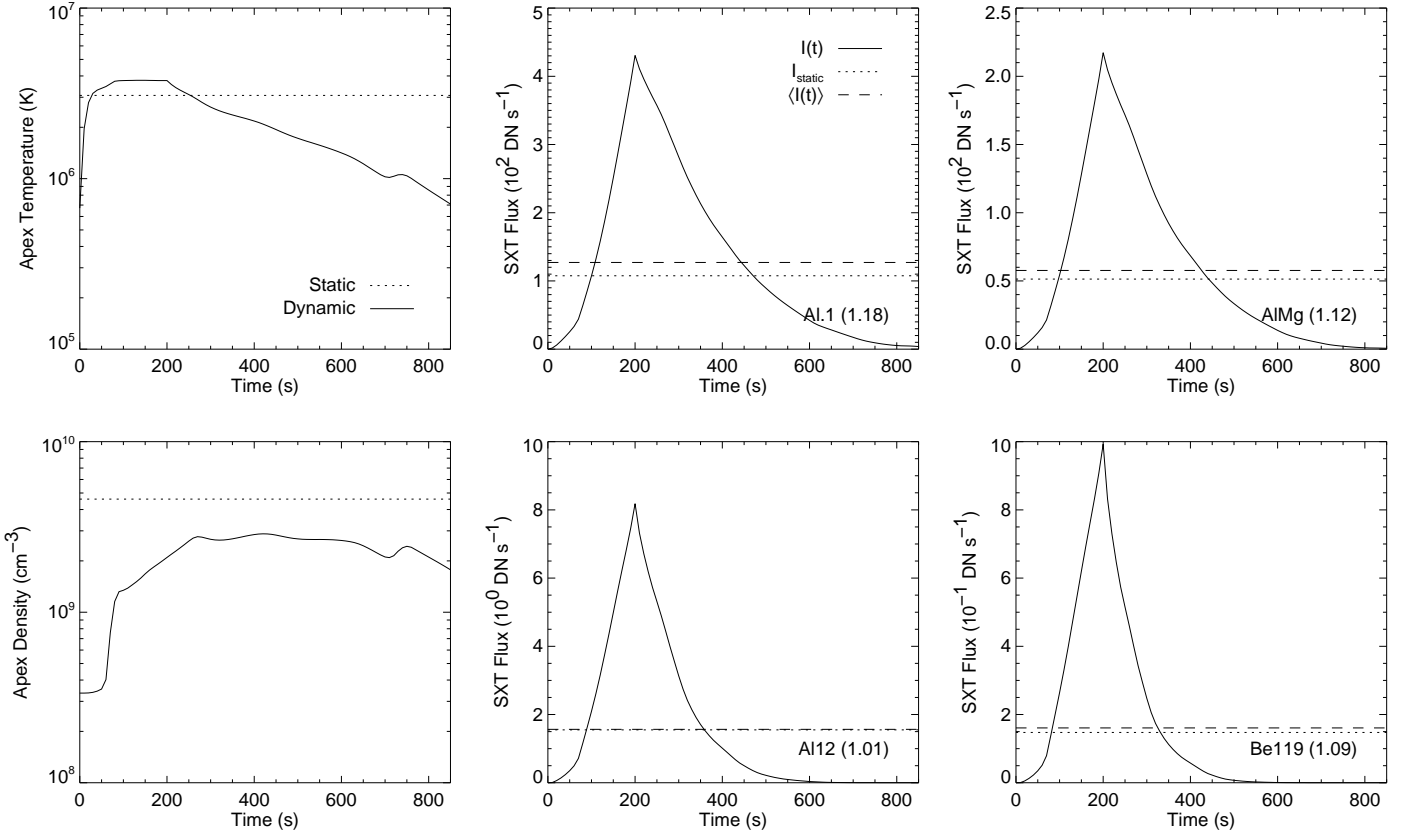}}
 \caption{A comparison of static and dynamic simulations of SXT
 emission. The parameters for the static case are $L=37$\,Mm and
 $\epsilon_S=0.0126$\,erg~cm$^{-3}$~s$^{-1}$, and are typical of the
 active region simulation. For the dynamic case $R=1.5$ and the
 heating duration is 200\,s. (\textit{right panels}) The evolution
 of the temperatures and densities averaged over the top 10\% of the
 loop apex. (\textit{right panels}) The evolution of the intensities in 4
 SXT filters. The SXT intensities from the static model, including the
 filling factor from the active region simulation, are also shown. For
 this case the SXT intensities from the time-averaged dynamic simulation
 are very close to those from the static simulation.}
 \label{fig:5}
 \end{figure*}

 An illustrative dynamical simulation is shown in
 Figure~\ref{fig:5}. Here $R=1.5$ and the heating duration is
 200\,s. For these parameters the apex densities are somewhat somewhat
 lower than the corresponding static solution. The time-averaging also
 reduces the SXT intensities significantly relative to their peak
 values. When the filling factor is included in the calculation of the
 SXT intensities from the static solutions, however, we see that the
 SXT intensities calculated from the two different simulations are
 very similar in all of the filters of interest. 

 Note that the computed intensities are somewhat dependent on the
 interval chosen for the time averaging. We assume that each small
 scale thread is heated once then allowed to cool fully before being
 heated again.  In practice, we terminate the dynamical simulation
 when the apex temperature falls below 0.7\,MK. The SOLFTM only has an
 adaptive mesh in the transition region and cannot resolve the
 formation of very cool material in the corona. Radiative losses
 become very large at low temperatures so the loops evolve very
 rapidly past this point and the time-averaged intensities should be
 only weakly dependent on when the dynamical simulation is stopped.

 While calculations such as this, which show that the SXT intensities
 computed from the dynamical simulation and those computed from the
 static simulation can be comparable, are encouraging, they represent
 a special case. In general, for a fixed value of $R$ the ratio of the
 dynamic and static intensities will be greater than 1 for shorter
 loops and smaller than 1 for longer loops. We would like to know what
 would happen if we performed dynamical simulations for all of the
 loops in AR8156.  Would the total intensities in the dynamical
 simulation match the observations? 

 The dynamical solutions we have investigated in this paper typically
 take about 500\,s to perform on an Intel Pentium 4-based
 workstation. For our 1956 field lines this amounts to about 11 days
 of cpu time. While such calculations can be done in principle,
 particularly on massively parallel machines with 100s of nodes,
 they're too lengthy for the exploratory work we consider here. To
 circumvent this computational limitation we consider a grid of
 solutions that encompass the range of loop lengths and heating rates
 that are present in our static simulation of this active region.

\begin{figure*}[t!]
 \centerline{%
 \includegraphics[scale=0.80,angle=90]{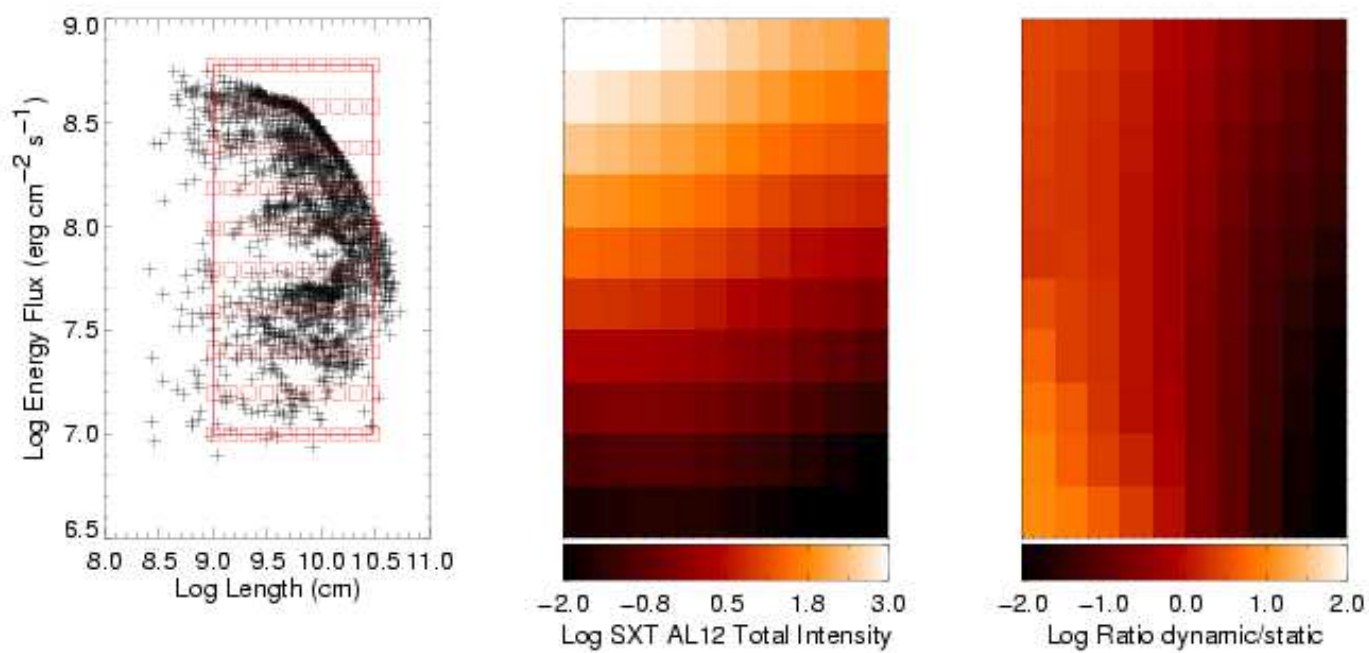}}
 \caption{(\textit{left panel}) A scatter plot of loop length ($L$)
 and energy flux ($\epsilon_SL$) for each field line in the static
 active region simulation. The red boxes indicate the values for which
 dynamic solutions have been calculated. The domain of the grid
 encompasses about 95\% of the total SXT intensity in the active
 region. (\textit{middle panel}) The total SXT Al12 intensity for each
 $L, \epsilon_S L$ pair in the grid for $R=1.5$. (\textit{right
 panel}) The ratio of the total SXT Al12 intensities in the dynamic
 and static simulations for $R=1.5$. The other SXT filters yield total
 intensities and ratios very similar to what is shown here for Al12.}
 \label{fig:6}
 \end{figure*}

 In Figure~\ref{fig:6} we also show a plot of total loop length ($L$)
 and energy flux ($\epsilon_S L$) for each field line in the static
 simulation of AR8156. Note that we use the energy flux instead of the
 volumetric heating rate because, as indicated in the plot, these
 variables are largely uncorrelated and we can use a simple
 rectangular grid. The volumetric heating rate and the loop length, in
 contrast, are correlated. The procedure we adopt is to calculate
 dynamical solutions for the $L, \epsilon_S L$ pairs on this grid,
 determine the total SXT intensities for these solutions, and then use
 interpolation to estimate the SXT intensities for each field line in
 the simulation. To investigate the effects of varying $R$ on the
 dynamical solutions we have computed $10\times10$ grids for $R=$1.0,
 1.25, 1.5, 1.75, and 2.0 for a total of 500 dynamical simulations.

 One important difference between these grid solutions and the static
 solutions discussed in the previous section is the loop geometry. In
 the static simulation the loop geometry is determined by the field
 line. In the dynamic simulation the loop is assumed to be
 perpendicular to the solar surface. Since the density scale height
 for high temperature loops is large this difference has only a small
 effect on the simulation of the SXT emission. The effect is much more
 pronounced for the lower temperature loops imaged in EIT and
 precludes the use of the interpolation grid for these intensities.

 In Figure~\ref{fig:6} we show the resulting SXT Al12 intensities for
 the $R=1.5$ grid of solutions. The most intense loops in the dynamic
 simulations are the shortest loops with the most intense
 heating. These loops come the closest to reaching equilibrium
 parameters with the finite duration heating. The faintest loops are
 the longest loops with the weakest heating. In Figure~\ref{fig:6} we
 also show the ratio of the SXT intensities from the dynamic and static
 simulations. That is, we compare the total intensity determined from
 the static solution with a heating rate of $\epsilon_S$ with the
 total intensity determined with the time dependent heating rate
 $\epsilon_D$ given in Equation~\ref{eq:heatingD}. These ratios
 indicate that for a significant region of this parameter space the
 total intensities in the dynamic heating are close to those computed
 for the static simulations. The shorter field lines are somewhat more
 intense while the longer field lines are generally fainter. This
 suggests that the dynamic intensities integrated over all of the
 field lines should produce results similar to the static
 simulation. As expected, for smaller values of $R$ these ratios are
 systematically smaller and for larger values of $R$ these ratios are
 systematically larger.

\begin{figure}[t!]
 \centerline{%
 \includegraphics[scale=0.75]{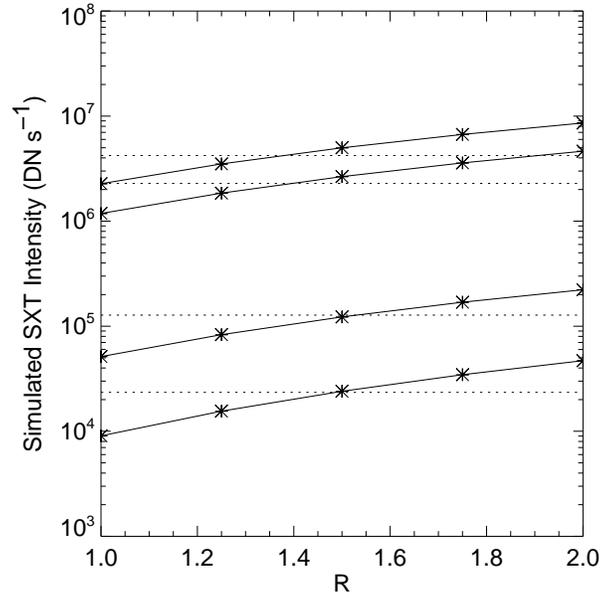}}
 \caption{Total SXT intensities from the dynamical simulations as a
 function of $R$, the ratio of impulsive to static heating rate. The
 dotted lines indicate the observed intensities. The simulation grid
 with $R=1.5$ best approximates the observed intensities.}
 \label{fig:7}
 \end{figure}

 We have used the results from all of the dynamic simulation grids to
 estimate the total SXT intensity in each filter as a function of
 $R$. The results are shown in Figure~\ref{fig:7}. For $R=1$ the total
 intensities are smaller than what is observed by about 50\%. For
 $R=2$ the intensities are all too large by about 100\%. For the
 $R=1.5$ case, which we have highlighted in Figures~\ref{fig:5} and
 \ref{fig:6}, the simulated total intensities are within 20\% of the
 measurements in all 4 filters, and this case come closest to
 reproducing the observations. The differences between the model
 calculations and the observations are not systematic. The calculated
 Al12 and Be119 intensities are very close to the observations while
 the Al1 and AlMg are a little too high. The duration of the heating,
 which we have chosen to be 200\,s, may explain this discrepancy, at
 least partially. If the heating were reduced in duration stronger
 heating (larger $R$) would be required to match the observed
 intensities. This would lead to higher temperatures and lead to
 somewhat different ratios among three filters.

\begin{figure*}[t!]
 \centerline{%
 \includegraphics[angle=90,scale=0.80]{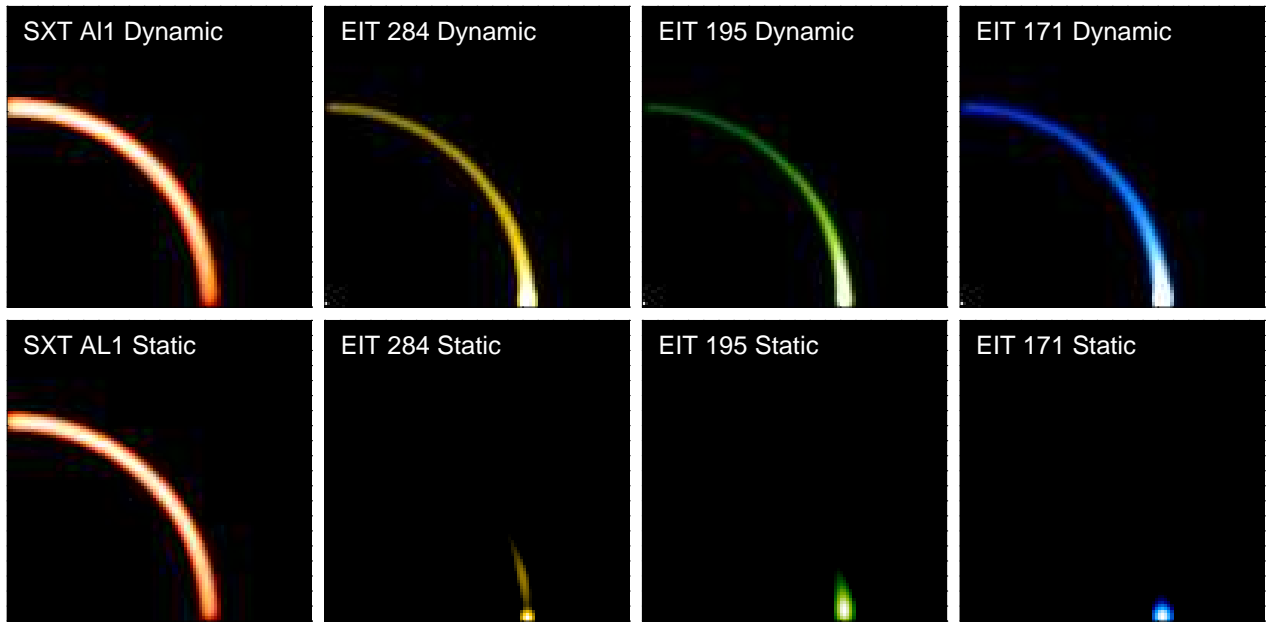}}
 \caption{The distribution of emission from dynamical and static
 heating models of the field line shown in
 \protect{Figure~\ref{fig:5}}. The same logarithmic scaling is used
 for each pair of images. The width of the point spread function has
 been chosen arbitrarily. The intensities shown for the dynamic case
 are time-averaged. The morphology of the loop imaged in SXT is very
 similar in both the dynamic and static heating scenarios. The dynamic
 heating gives rise to a much more even distribution of intensity
 along the field line at the cooler temperatures.}
 \label{fig:8}
 \end{figure*}

 One of the primary motivations for introducing dynamic modeling is
 the inability of static models to account for the EUV observations at
 lower temperatures. Because we have used interpolation grids to infer
 the intensities it is not possible to compute images similar to those
 presented in Figure~\ref{fig:4}. We can, however, consider the
 morphology of individual loops with the static and dynamic heating
 scenarios. To investigate this we calculate the time-averaged
 intensity in SXT and EIT along the loop. As illustrated in
 Figure~\ref{fig:8}, the dynamic heating is clearly moving in the
 right direction. The morphology of the high temperature plasma imaged
 with SXT is largely unchanged in the dynamic simulation while the EUV
 emission shows full loops. This suggests that the dynamic simulations
 of active regions will look much closer to the observations shown in
 Figure~\ref{fig:2} than the synthetic images calculated from static
 heating models.

 One implication of Figure~\ref{fig:8} is that relatively cool loops
 imaged in the EUV should be co-spatial with high temperature loops
 imaged in soft X-rays. There is some evidence that this is not
 observed.  \cite{schmieder2004} and \cite{nitta2000}, for example,
 argue that the EUV loops may be observed near soft X-ray loops, but
 that they are generally not co-spatial. It is possible, however, that
 the geometry of the loops changes as they cool
 \citep{winebarger2005}. \cite{antiochos2003} suggest that the
 observed EUV loop emission in an active region is not bright enough
 to account for the cooling of soft X-ray loops. However, the contrast
 between the background corona and the EUV loops is low
 \citep{cirtain2006}, and it is possible that the total EUV intensity
 in an active region is consistent with the cooling of high
 temperature loops. 

 \section{Summary and Discussion}   

 We have investigated the use of static and dynamic heating models in
 the simulation of AR8156. Recent work has shown that static models
 can capture many of the observed properties of the high temperature
 soft X-ray emission from solar active regions and our results confirm
 this. We are able to reproduce both the total intensity and the
 general morphology of this active region with a static heating
 model. Furthermore, our results show that this agreement extends to
 the the hotter SXT filters which have not been considered before.

 The application of dynamic heating models to active region emission
 on this scale has not been considered previously. Only the properties
 of individual loops have been explored \citep[e.g.][]{warren2002b}. The
 computational complexity of the dynamical simulations precludes the
 calculation of individual solutions for each field line and we have
 utilized interpolation grids for estimating the expected SXT
 intensities for each field line. We find that it is possible to
 reproduce the observed SXT intensities in 4 filters, including the
 high temperature Al12 and Be119 filters, using the dynamical model.

 Conceptually, the simple dynamical heating model investigated here,
 where we assume that the emission from a solar feature results from
 the superposition of many, very fine structures that are in various
 stages of heating and cooling, is closely related to the nanoflare
 model of coronal heating \cite[e.g.,][]{parker1983,cargill1994}. The
 use of time-averaged intensities computed from the dynamical
 simulations implicitly assumes that the heating is very coherent,
 with each infinitesimal thread being heated once and then allowed to
 cool and drain before being heated again. Other scenarios are
 possible, such as heating events on larger scale threads that are
 distributed randomly in time. The spatial and temporal
 characteristics of coronal heating is likely to fall somewhere in
 between these extremes. The analysis of high spatial resolution
 (0.5\arcsec) EUV images suggests that current solar instrumentation
 may be close to resolving individual threads in the corona
 \citep{aschwanden2005a,aschwanden2005b}, but considerable work
 remains to be done to determine the fundamental spatial scale for
 coronal heating.

 The geometrical properties of coronal threads is also unclear at
 present. We have assumed constant cross sections in our modeling,
 consistent with the observational results
 \citep{klimchuk1992,watko2000}.  In the static modeling of solar
 active regions there has been some evidence that the loops with
 expanding cross sections better reproduce the observations
 \citep{schrijver2004,warren2006b}. Detailed comparisons between
 simulated and observed solar images are needed to resolve this
 issue. 
  
 Despite the many limitations to our modeling the results that we have
 presented are encouraging and provide a framework for further
 exploration. The highest priority for future work is the full
 dynamical simulation of solar active regions without the use of
 interpolation grids so that synthetic soft X-ray and EUV images can
 be computed and compared with observations. Another priority is the
 comparison of active region simulations with spatially and spectrally
 resolved observations from the upcoming Solar-B mission. Spectral
 diagnostics, such as Doppler velocities and nonthermal widths, are
 another dimension that have not been explored in the context of this
 modeling. 
 
%% ----------------------------------------------------------------------
%% --- ACKNOWLEDGMENTS --------------------------------------------------
%% ----------------------------------------------------------------------

\acknowledgments The authors would like to thank John Mariska for his
helpful comments on the manuscript. Yohkoh is a mission of the
Institute of Space and Astronautical Sciences (Japan), with
participation from the U.S. and U.K. The EIT data are courtesy of the
EIT consortium. This research was supported by NASA's Supporting
Research and Technology and Guest Investigator programs and by the
Office of Naval Research.

%% ----------------------------------------------------------------------
%% --- REFERENCES -------------------------------------------------------
%% ----------------------------------------------------------------------

%% ----------------------------------------------------------------------
%% --- FIGURES ----------------------------------------------------------
%% ----------------------------------------------------------------------

\clearpage

\clearpage

\clearpage

\clearpage

\clearpage

\clearpage

\clearpage

\clearpage

\end{document}